\documentclass{esannV2}
\usepackage[dvips]{graphicx}
\usepackage[latin1]{inputenc}
\usepackage{amssymb,amsmath,array}
\usepackage{subfigure}

\begin{document}

\providecommand{\norm}[1]{\lVert#1\rVert}

\renewcommand{\figurename}{Fig.\ }

% make bold
\newcommand{\bd}[1]{\mbox{\boldmath $#1$}}
\newcommand{\RR}[1] {\mathit{R}^{#1}} 
\newcommand{\kronDelta}[2]{\delta_{#1,#2}}
\newcommand{\dx}[2] {\frac{\partial #2} {\partial x_{#1}}}
% second order partial derivative
\newcommand{\dxdx}[3] {\frac{\partial^{2}#3} {\partial x_{#1} x_{#2}}}
% superscript indicator
\newcommand{\Nth}[1]{^{(#1)}}
% subscript indicator
\newcommand{\nth}[1]{_{(#1)}}
\newcommand{\vect} {\bd{t}}
\newcommand{\vectN}[1] {\bd{t}\Nth{#1}}
\newcommand{\GaussianPDF}[3] { {\mathcal N}(#1;#2,#3)}
\newcommand{\KLD}[2] {D_{KL}\lbrack #1 \vert\vert #2 \rbrack}
\newcommand{\argmax}[2] {\mathop{argmax}_{#1}\bigg\{#2\bigg\}}
\newcommand{\argmin}[2] {\mathop{argmin}_{#1}\bigg\{#2\bigg\}}
\newcommand{\expect}[1] {E\lbrack {#1} \rbrack}
\newcommand{\expectWRT}[2] {E_{#1} \lbrack {#2} \rbrack}

%\newcommand{\thetanew}{\bd{\theta}}
%\newcommand{\thetaold}{\bd{\theta}^\prime}
%%%%%%%%%%%%% COMMANDS FOR GTM %%%%%%%%%%%%%%%%
\newcommand{\latent}[1]{\bd{x}_{#1}}
\newcommand{\model}[1]{\mbox{$p(\cdot|#1)$}}
%============================================================
% NEURAL NETWORKS
%============================================================

%============================================================
% HIDDEN MARKOV MODELS AND HIDDEN MARKOV TREE MODELS
%============================================================
\newcommand{\Seq} {\bd{S}}
\newcommand{\seq} {\bd{s}}
\newcommand{\SeqN}[1] {\bd{S}\Nth{#1}}
\newcommand{\seqN}[1] {\bd{s}\Nth{#1}}

\newcommand{\Symb}[1] {\bd{S}_{#1}}
\newcommand{\symb}[1] {\bd{s}_{#1}}
\newcommand{\SymbN}[2] {\bd{S}_{#1}\Nth{#2}}
\newcommand{\symbN}[2] {\bd{s}_{#1}\Nth{#2}}

\newcommand{\Zcomp}[2] {z^{(#1)}_{#2}}
\newcommand{\Zstate}[3] {z^{(#1,#2)}_{#3}}
\newcommand{\Ztrans}[4] {z^{(#1,#2)}_{#3 \rightarrow #4}}
\newcommand{\State}[1]{Q_{#1}}
\newcommand{\state}[1]{q_{#1}}

\newcommand{\Tree}{\bd{Y}}
\newcommand{\tree}{\bd{y}}
\newcommand{\TreeN}[1]{\bd{Y}\Nth{#1}}
\newcommand{\treeN}[1]{\bd{y}\Nth{#1}}

\newcommand{\Subtree}[1]{\bd{Y}_{#1}}
\newcommand{\subtree}[1]{\bd{y}_{#1}}
\newcommand{\SubtreeN}[2]{\bd{Y}_{#1}\Nth{#2}}
\newcommand{\subtreeN}[2]{\bd{y}_{#1}\Nth{#2}}

\newcommand{\parent}[1]{\rho{(#1)}}
\newcommand{\child}[1]{c{(#1)}}

\newcommand{\NodeLabel}[1]{\bd{O}_{#1}}
\newcommand{\nodeLabel}[1]{\bd{o}_{#1}}
\newcommand{\NodeLabelN}[2]{\bd{O}_{#1}\Nth{#2}}
\newcommand{\nodeLabelN}[2]{\bd{o}_{#1}\Nth{#2}}

\newcommand{\SubTreeExcl}[2]{\bd{Y}_{#1 \backslash #2}}
\newcommand{\subTreeExcl}[2]{\bd{y}_{#1 \backslash #2}}
\newcommand{\SubTreeExclN}[3]{\bd{Y}_{#1 \backslash #2}\Nth{#3}}
\newcommand{\subTreeExclN}[3]{\bd{y}_{#1 \backslash #2}\Nth{#3}}

%----------- MATRICES FOR  HIDDEN MARKOV TREES -------------
\newcommand{\Ainit}{\bd{A}^{(\bd{\pi})}}
\newcommand{\Atrans}[1]{\bd{A}^{(\bd{T}_{#1})}}
\newcommand{\Aemiss}[1]{\bd{A}^{(\bd{B}_{#1})}}

%style file for ESANN manuscripts
\title{Autoencoding Time Series for Visualisation}

%***********************************************************************
% AUTHORS INFORMATION AREA
%***********************************************************************
\author{Nikolaos Gianniotis$^1$, Dennis K\"{u}gler$^1$, Peter Ti\v{n}o$^2$, Kai Polsterer$^1$
and Ranjeev Misra$^3$
%
% Optional short acknowledgment: remove next line if non-needed
%\thanks{This is an optional funding source acknowledgement.}
%
% DO NOT MODIFY THE FOLLOWING '\vspace' ARGUMENT
\vspace{.3cm}\\
1- Astroinformatics - Heidelberg Institute of Theoretical Studies \\
 Schloss-Wolfsbrunnenweg 35 D-69118 Heidelberg - Germany
\vspace{.1cm}\\
2 - School of Computer Science - The University of Birmingham \\
Birmingham B15 2TT - UK
\vspace{.1cm}\\
3 - Inter-University Center for Astronomy and Astrophysics \\
Post Bag 4, Ganeshkhind, Pune-411007 - India
}
%***********************************************************************
% END OF AUTHORS INFORMATION AREA
%***********************************************************************

\maketitle

\begin{abstract}
We present an algorithm for the visualisation of time series. To that end we employ echo state networks to convert time series into a suitable vector representation which is capable of capturing the latent dynamics of the time series. Subsequently, the obtained vector representations are put through an autoencoder and the visualisation is constructed using the activations of the ``bottleneck". The crux of the work lies with defining an objective function 
that quantifies the reconstruction error of these representations in a principled manner.
We demonstrate the method on synthetic and real data.
\end{abstract}

%=========================================================================
\section{Introduction}
%=========================================================================

Time series are often considered a challenging data type to handle in 
machine learning tasks. Their variable-length nature has forced
the derivation of feature vectors that capture various characteristics, e.g. \cite{Richards2011}. However,
it is unclear how well such (often handcrafted)
features  express the potentially complex latent dynamics of time series. 
Time series exhibit long-term dependencies
which
 must be taken into account when comparing two time series for similarity.
This temporal nature makes the use of common designs, e.g. RBF kernels,
problematic. Hence, more attentive algorithmic designs are needed and indeed in classification scenarios there have been works \cite{Jaakkola1998, Jebara2004, Chen2013} that successfully account for the particular nature of time series.

In this work we are interested in visualising time series.
We propose a fixed-length vector representation for representing
sequences that is based on the Echo State Network (ESN) \cite{Jaeger2001} architecture.
The great advantage of ESNs is the fact that the hidden part, the reservoir of nodes,
is fixed and only the readout weights need to be trained. 
%The readout weights interact linearly
%with the reservoir and can be computed by solving a (regularised) least-squares problem.
In this work, we take the view that the readout weight vector 
is a good and comprehensive representation for a time series.
%Hence, we convert each time series into a fixed-length vector representation.

In a second stage, we employ an autoencoder \cite{Kramer1991} 
that reduces the dimensionality of the readout weight vectors.
However, employing the usual $L_2$ objective function for measuring reconstruction
is inappropriate.
What we are really interested in is not how well the readout weight vectors are reconstructed
in the $L_2$ sense, but how well each reconstructed readout weight vector
can still reproduce its respective time series when plugged back to the \textit{same, fixed} ESN reservoir.
To that end, we introduce a more suitable objective function for measuring the reconstruction
quality of the autoencoder.

%=========================================================================
\section{Echo state network cost function}
%=========================================================================

An ESN is a recurrent discrete-time neural network.
It processes time series composed by a sequence of observations which we denote
by\footnote{For brevity we assume univariate time series, i.e. $y(t)\in\RR{}$. }  $\bd{y}=(y(1), y(2), \dots, y(T))$.
The task of the ESN is given  $y(t)$ as an input to predict  $y(t+1)$.
An ESN is typically formulated using the following two equations:
\begin{eqnarray}
\bd{x}(t+1) &=& h( \bd{u} \bd{x}(t) + \bd{v} y(t)) \ , \label{eq:reservoir}\\
y(t+1) &=& \bd{w} \bd{x}(t+1)\  , \label{eq:readout}
\end{eqnarray}
where $\bd{v}\in\RR{N\times 1}$ is the input weight, $\bd{x}(t)=[x_1,\dots,x_N]\in\RR{N\times 1}$ are the latent state activations of the reservoir,  $\bd{u}\in\RR{N\times N}$ the weights of the reservoir units, $\bd{w}\in\RR{N\times 1}$ the readout 
weights\footnote{Bias terms can be subsumed into weight vectors \bd{v} and \bd{u} but are ignored here for brevity.}. $N$ is the number of hidden reservoir units.
Function $h(\cdot)$ is a nonlinear function, e.g. $\tanh$,
applied element-wise. According to ESN methodology \cite{Jaeger2001}
parameters $\bd{v}$ and $\bd{u}$ in  Eq.~\eqref{eq:reservoir} are randomly generated and fixed. The only trainable parameters are the readout weights $\bd{w}$ in Eq.~\eqref{eq:readout}.  

Training involves feeding at each time step $t$ an observation $y(t)$  and recording the resulting activations $\bd{x}(t)$
row-wise into a matrix $\bd{X}$. Typically, some initial observations are dismissed 
in order to  ``washout" \cite{Jaeger2001} the initial arbitrary reservoir state.
The following objective function $\ell$ is minimised:
\begin{eqnarray}
\ell(\bd{w}) = \frac{1}{2}\|\bd{X}\bd{w} - \bd{y}\|^2  + \frac{1}{2}\lambda^2 \|\bd{w}\|^2 \ ,
\label{eq:ESN_objective}
\end{eqnarray}
where $\lambda$ is a regularisation term.
How well vector $\bd{w}$ models $\bd{y}$ with respect to the fixed reservoir is measured by objective $\ell$.
The optimal solution is  $\bd{w} = (\bd{X}^T \bd{X}+ \lambda^2 \bd{I})^{-1} \bd{X}^T  \bd{y} $
where $\bd{I}$ is the identity matrix. We express this as a function $g$ that maps time series to readout weights:
\begin{eqnarray}
g(\bd{y}) = (\bd{X}^T \bd{X} + \lambda^2 \bd{I})^{-1} \bd{X}^T  \bd{y} = \bd{w} \ .
\label{eq:esn_project}
\end{eqnarray}

%=========================================================================
\section{Vector representation for time series}
%=========================================================================

Given a fixed ESN reservoir, for each time series in the dataset
we determine its best readout weight vector and take it to be
its new representation \textit{with respect to this reservoir}.

%-------------------------------------------------
\subsection{ESN reservoir construction}
%-------------------------------------------------

Typically, parameters $\bd{v}$ and $\bd{u}$ in Eq.~\eqref{eq:reservoir} are set stochastically \cite{Jaeger2001}.
To eliminate dependence on random initial conditions when constructing the ESN reservoir,
we strictly follow the deterministic scheme\footnote{We stress that our algorithm is not dependent on this deterministic scheme for constructing ESNs; in fact it also works with the ``standard"
stochastically constructed ESN type as in \cite{Jaeger2001}.} in \cite{Rodan2011}.
Accordingly, we fix the topology of the reservoir by organising the reservoir units in a cycle
using the \textit{same} coupling weight $a$. Similarly, all elements in $\bd{v}$ are assigned the
same absolute value $b>0$ with signs determined by an aperiodic sequence as specified in \cite{Rodan2011}.
Further, following this methodology we determine values for  $a$ and $b$ by cross-validation.
The  combination $a,b$ with the lowest test error is used to instantiate
the ESN reservoir that subsequently encodes the time series as readout weights.

%-------------------------------------------------
\subsection{Encoding time series as readout weights}
%-------------------------------------------------

Given the fixed reservoir, specified by $a$ and $b$, we encode each
time series $\bd{y}_j$ in the dataset by the readout weights $\bd{w}_j$
using function $g(\bd{y}_j)=\bd{w}_j$ (see Eq.~\eqref{eq:esn_project}).
We emphasise that \textit{all} time series $\bd{y}_j$
are encoded with respect to the \textit{same} fixed reservoir.
Hence dataset $\{\bd{y}_1, \dots, \bd{y}_J\}$ is now replaced by
 %$\{g(\bd{y}_1), \dots, g(\bd{y}_J)\} =%
$\{\bd{w}_1, \dots, \bd{w}_J\}$.

%=========================================================================
\section{Autoencoding with respect to the fixed reservoir}
%=========================================================================

The autoencoder \cite{Kramer1991} learns 
an identity mapping by training on targets identical to the inputs.
Learning is restricted by the bottleneck that forces
the autoencoder to reduce the dimensionality of the inputs, and hence
the output  is only an approximate reconstruction of the input.
By setting the number of neurons in the bottleneck to two, the
bottleneck activations can be interpreted as two-dimensional projection coordinates $\bd{z}\in\RR{2}$
and used for visualisation.

The autoencoder is the composition of an encoding $f_{enc}$ and a decoding 
$f_{dec}$ function. Encoding maps inputs to coordinates, $f_{enc}(\bd{w})= \bd{z}$, while decoding approximately maps coordinates back to inputs, 
$f_{dec}(\bd{z}) = \bd{\tilde{w}}$.
The complete autoencoder is a function 
$f(\bd{w};\bd{\theta})=f_{dec}(f_{enc}(\bd{w})) = \bd{\tilde{w}}$,
where  $\bd{\theta}$ are the weights of the autoencoder trained by backpropagation.

%-------------------------------------------------
\subsection{Training mode}
%-------------------------------------------------

Typically, training the autoencoder involves minimising the $L^2$ norm between inputs and reconstructions 
over the weights $\bd{\theta}$:
\begin{eqnarray}
\frac{1}{2} \sum_{j=1}^{J} \| f(\bd{w}_j;\bd{\theta}) - \bd{w}_j\|^2  .
\label{eq:L2_AE_objective}
\end{eqnarray}
However, this objective measures only how good the reconstructions $\bd{\tilde{w}}_j$ are in the $L_2$ sense.
What we are really interested in is how well the reconstructed weights $\bd{\tilde{w}}_j$ are still
a good readout weight vector when plugged back to the fixed reservoir.
This is actually what the objective function $\ell$ in Eq.~\eqref{eq:ESN_objective} measures.
This calls for a modification in the objective function Eq.~\eqref{eq:L2_AE_objective} of the autoencoder:
\begin{eqnarray}
\frac{1}{2}  \sum_{j=1}^{J}  \ell_j(f(\bd{w}_j;\bd{\theta})) = \frac{1}{2} \sum_{j=1}^{J}  \| \bd{X}_j f(\bd{w}_j;\bd{\theta}) - \bd{y}_j \|^2 + \frac{1}{2}\lambda^2 \|f(\bd{w}_j;\bd{\theta})\|^2 \ ,
\label{eq:AE_objective}
\end{eqnarray}
where $\ell_j$  and $\bd{X}_j$ are the objective function
and hidden state activations associated with time series $\bd{y}_j$  (see Eq.~\eqref{eq:ESN_objective}).
The weights $\bd{\theta}$ of the autoencoder can now be  trained via backpropagation using the modified objective   in Eq.~\eqref{eq:AE_objective}.

%-------------------------------------------------
\subsection{Testing mode}
%-------------------------------------------------

Having trained the autoencoder $f$ via backpropagation, we would like to project new incoming time series $\bd{y}^*$ to coordinates $\bd{z}^*$.
To that end we first use the fixed ESN reservoir to encode the time series as a readout weight vector $g(\bd{y}^*)=\bd{w}^*$ (see Eq.~\eqref{eq:esn_project}).
The readout weight vector $\bd{w}^*$ can then be projected using the encoding part of the autoencoder
to obtain the projection $f_{enc}(\bd{w}^*)= \bd{z}^*$.

%=========================================================================
\section{Experiments and Results}
%=========================================================================

\begin{figure}
\centering
\subfigure{\includegraphics[bb=95 338 492 499,clip=true,width=5.5cm]{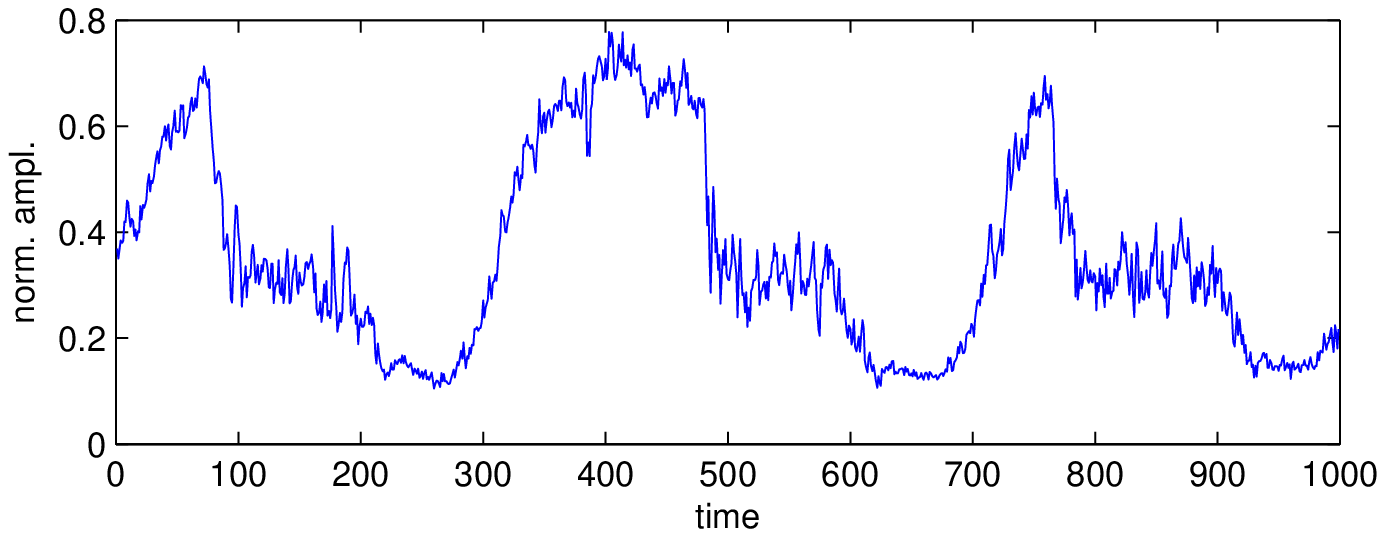}}
\subfigure{\includegraphics[bb=95 338 492 499,clip=true,width=5.5cm]{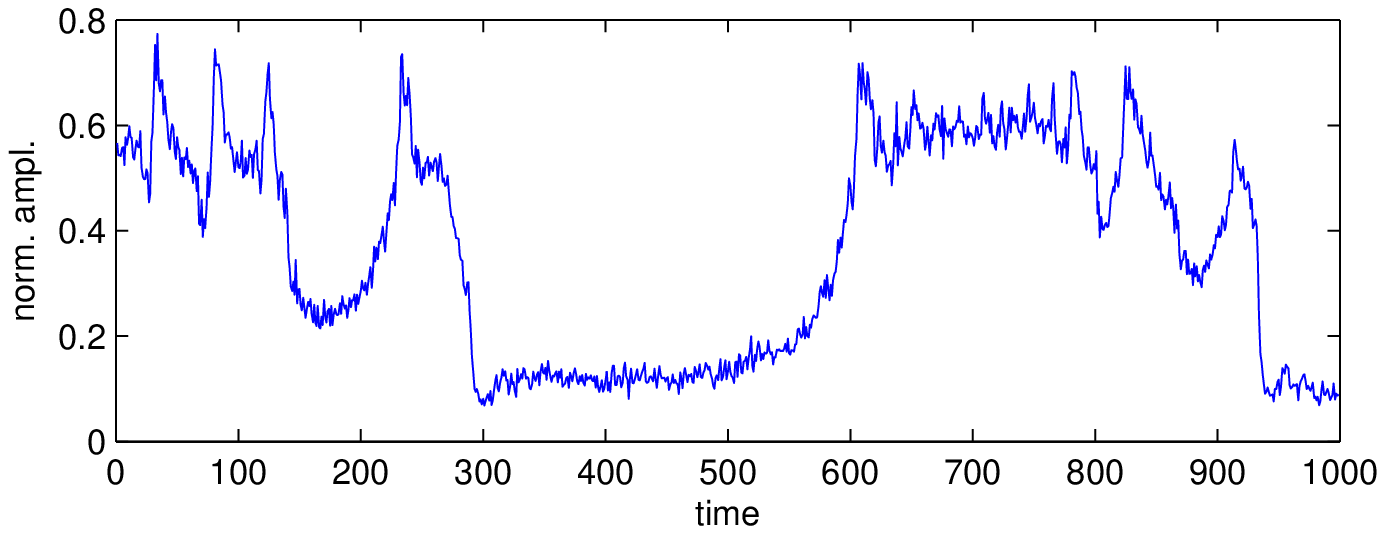}}
\caption{Example X-ray radiation regimes $\beta$ (left) and $\kappa$ (right).}
\label{fig:example_regimes}
\end{figure}

\begin{figure}[h!]
\centering
\subfigure[NARMA by t-SNE.]{\includegraphics[bb=87 214 517 627, clip=true, width=4.8cm]{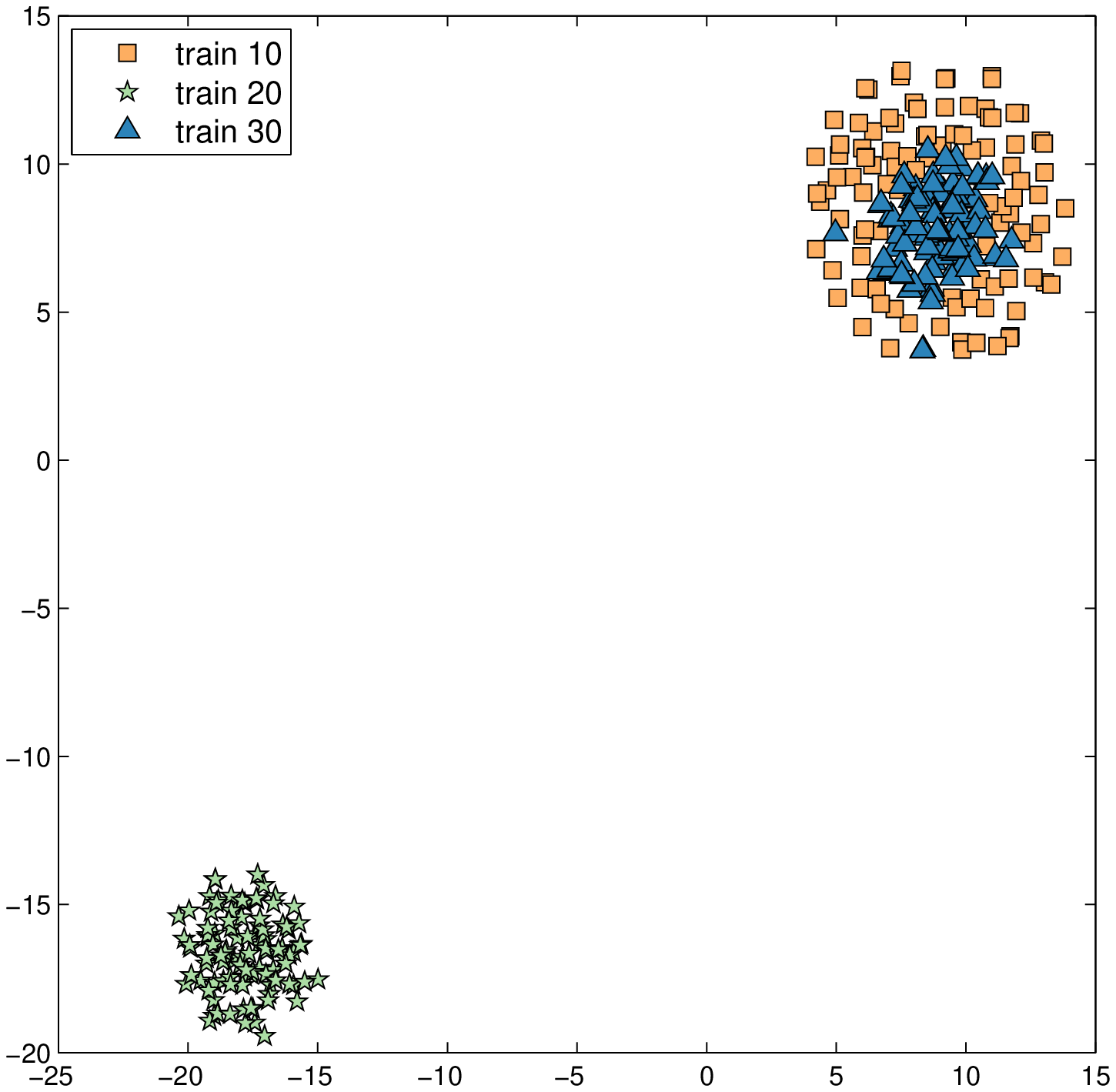}}
\subfigure[NARMA by our method.]{\includegraphics[bb=87 214 517 627, clip=true, width=4.8cm]{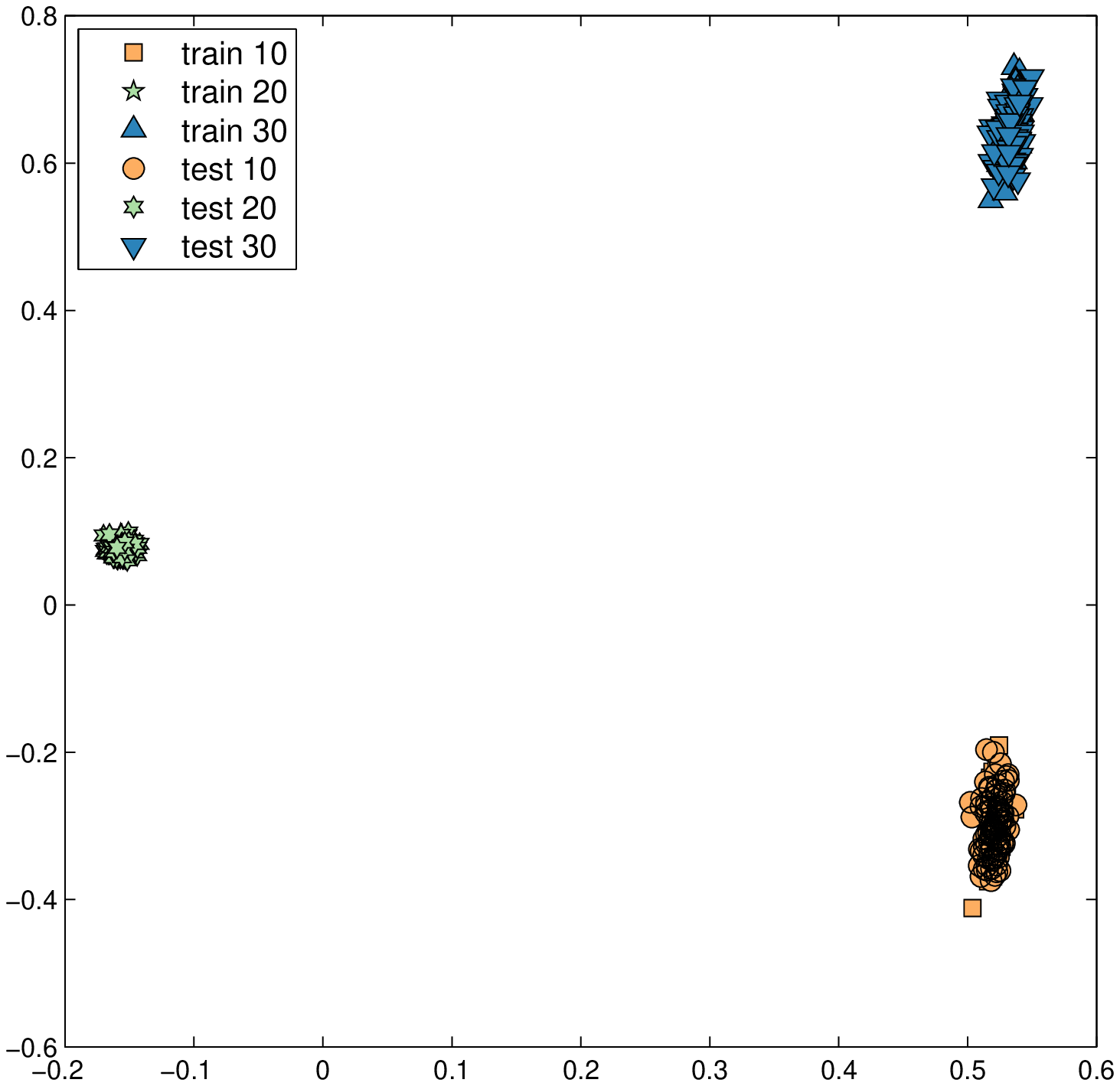}}
\subfigure[Cauchy by t-SNE.]{\includegraphics[bb=87 214 517 627, clip=true, width=4.8cm]{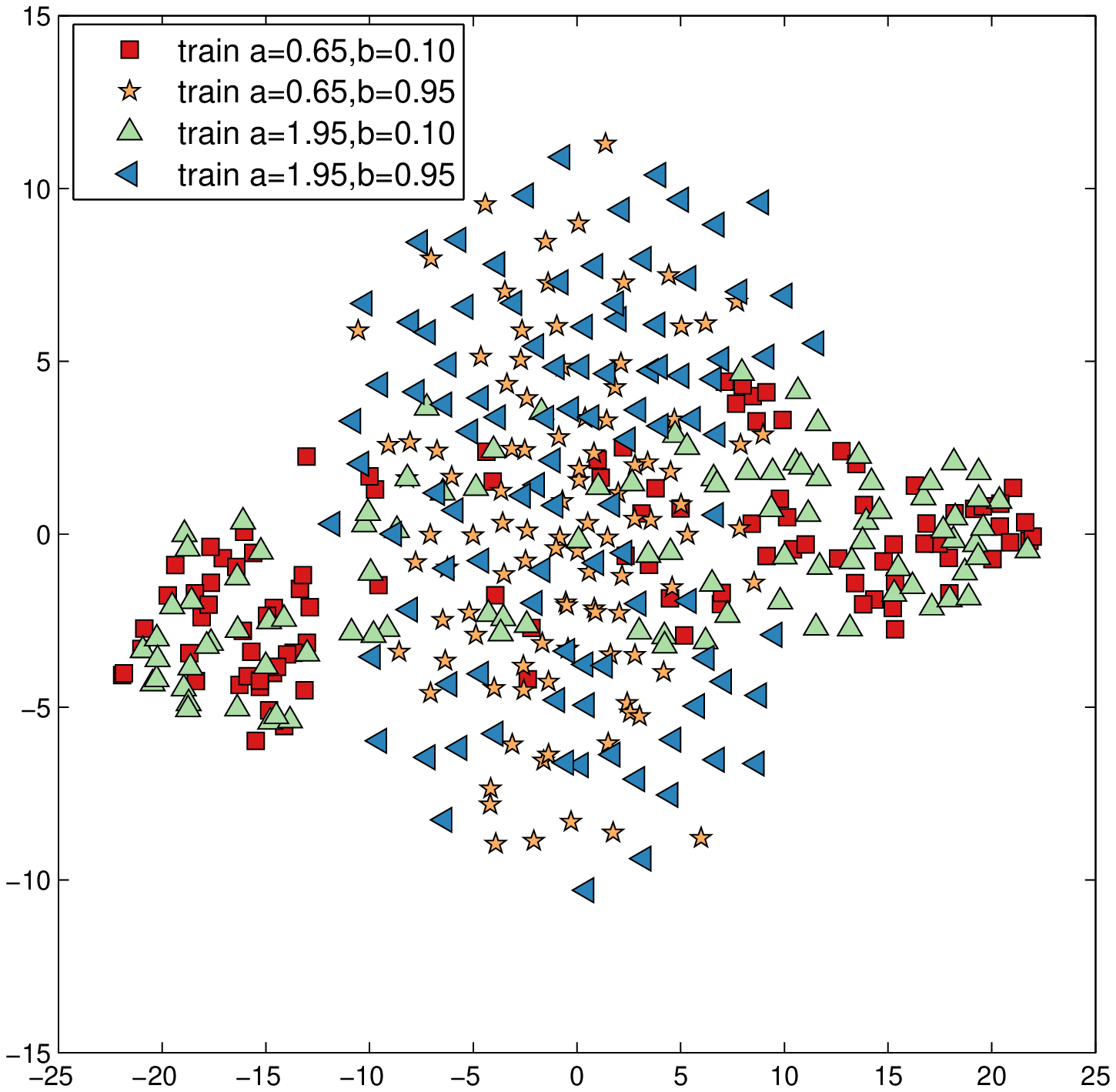}}
\subfigure[Cauchy by our method.]{\includegraphics[bb=87 214 517 627, clip=true, width=4.8cm]{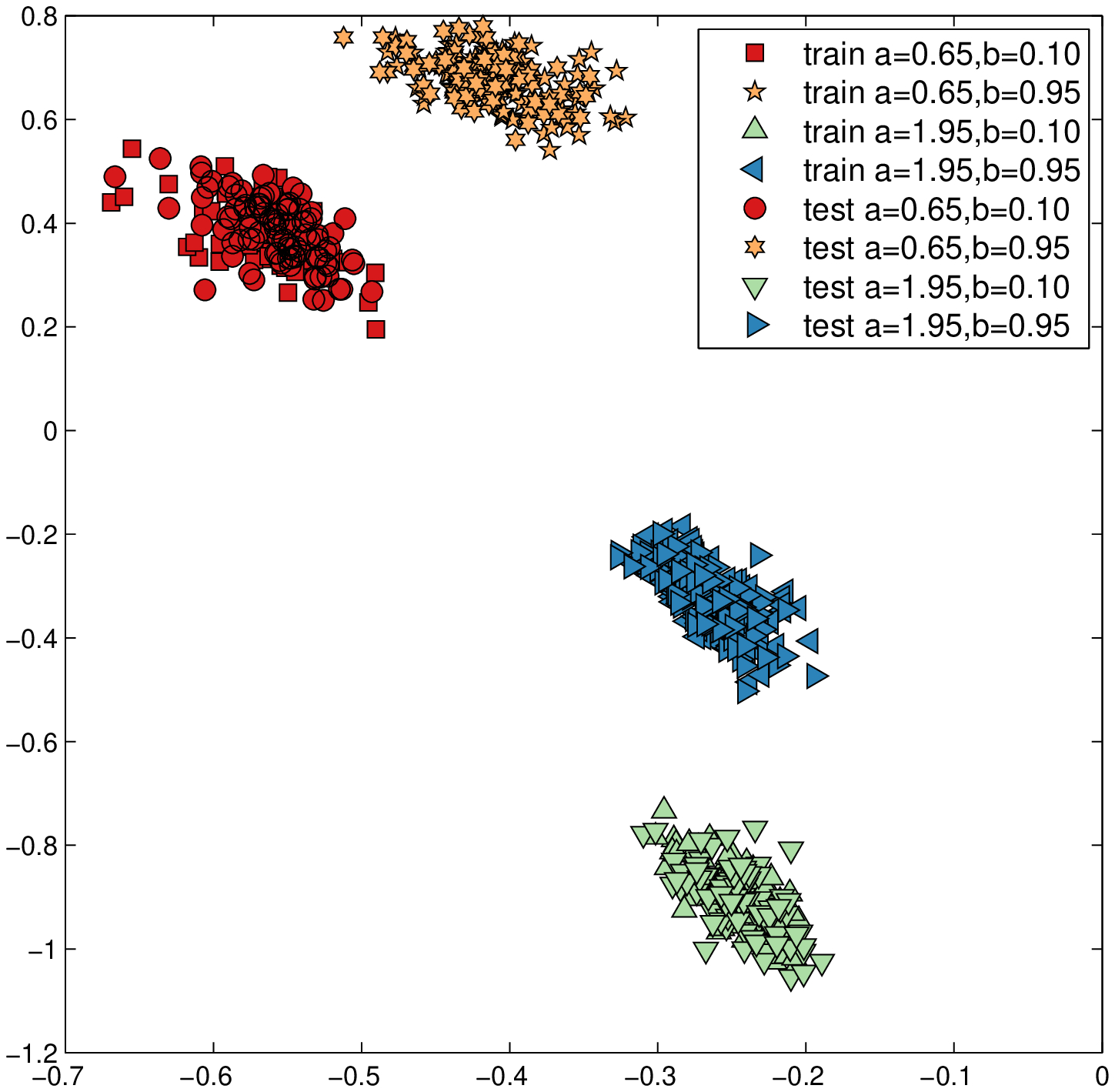}}
\subfigure[X-ray radiation by t-SNE.]{\includegraphics[bb=87 214 517 627, clip=true, width=4.8cm]{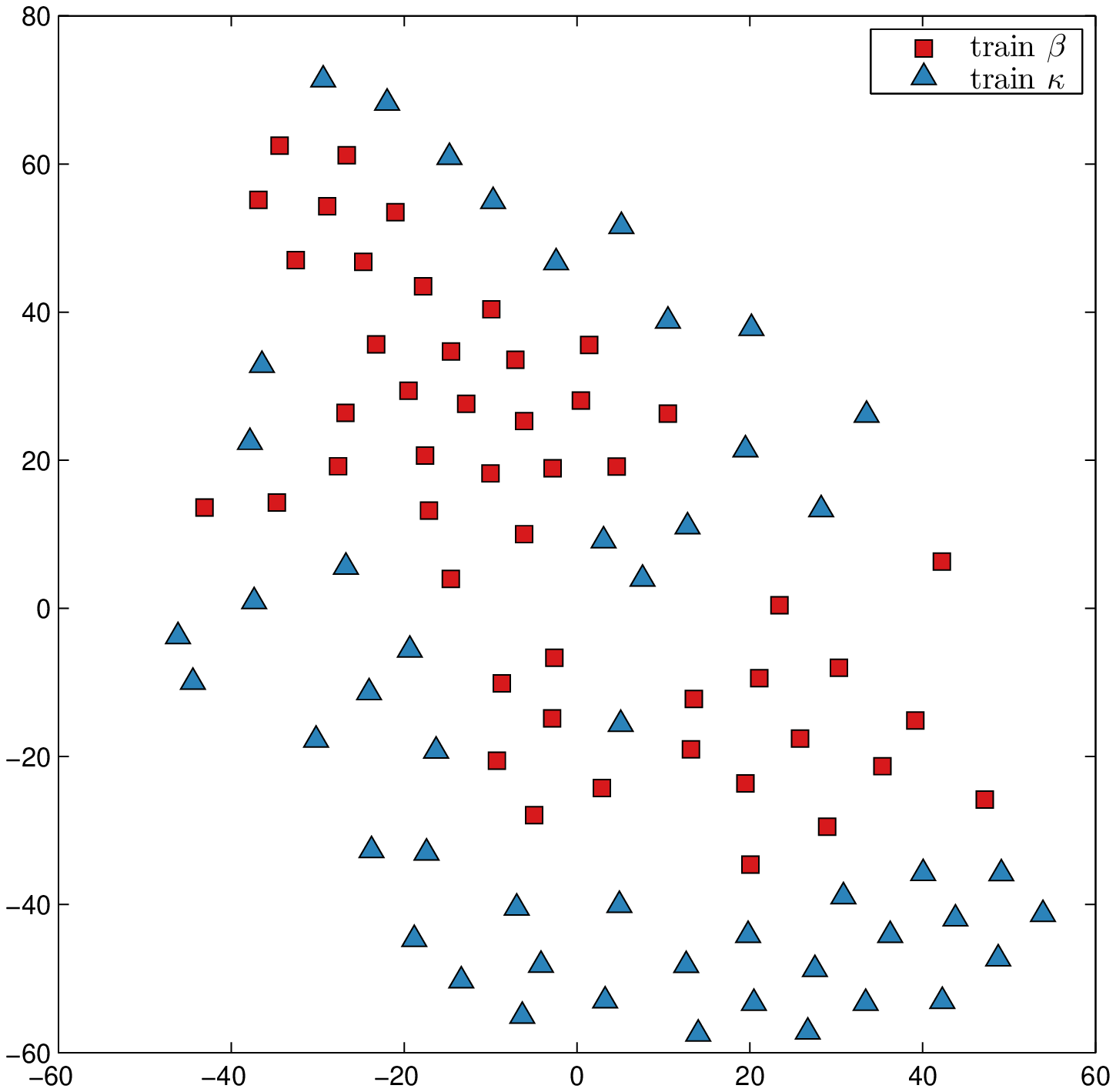}}
\subfigure[X-ray radiation by our method.]{\includegraphics[bb=87 214 517 627, clip=true, width=4.8cm]{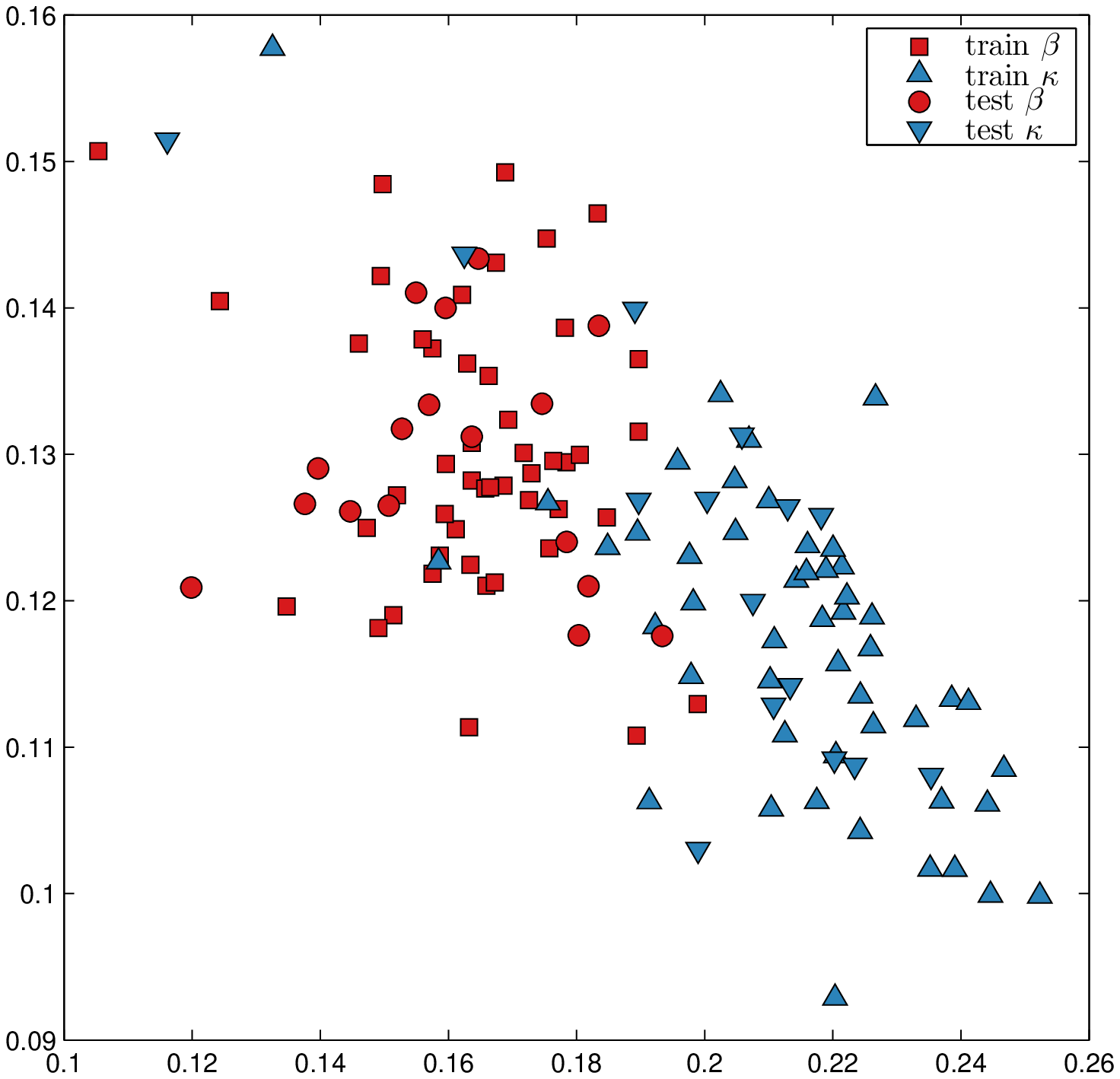}}
\caption{Colours represent classes. The proposed algorithm supports out-of-sample visualisation, hence markers $\bullet$ and  $\blacktriangle$ are the projections of the training and testing data respectively.
Note that in the NARMA and Cauchy plots $\bullet$ and  $\blacktriangle$ heavily overlap.}
\label{fig:results}
\end{figure}

We present results on two synthetic datasets and on a real astronomical dataset.
In all experiments we constructed the ESN reservoir deterministically according 
to \cite{Rodan2011}
and fixed the size of the reservoir to $N=200$. We used a washout period of $200$ observations.
Regularisation parameter $\lambda$ for the ESNs was  fixed to $10^{-4}$. 
The number of neurons in the hidden layers of the autoencoder was set to $10$.
The proposed algorithm can handle out-of-sample data and hence apart from 
projecting training data only, we also project unseen test data.
We apply no normalisation to the datasets.
Moreover, we also constructed visualisations using the popular t-SNE algorithm \cite{Maaten2008} on the raw signals. We found the visualisation produced by t-SNE did not differ greatly over a range of perplexities $\{5,10,\dots,50\}$.

%%------------------------------------------------------------------
\vspace{0.2cm}
%\newline
\noindent \textbf{NARMA:} We generated sequences from the following
NARMA classes \cite{Rodan2011} of order $10, 20, 30$, of length $800$, using the following equations respectively:
\begin{eqnarray}
y(t+1) \!\!\!\! &=& \!\!\!\! 0.3y(t) + 0.05y(t)\sum_{i=0}^{9} y(t-i) + 1.5s(t-9)s(t) + 0.1, \notag\\
y(t+1) \!\!\!\! &=& \!\!\!\! \tanh(0.3y(t) + 0.05y(t)\sum_{i=0}^{19} y(t-i) + 1.5s(t-19)s(t) + 0.01) + 0.2, \notag\\
y(t+1) \!\!\!\! &=& \!\!\!\! 0.2y(t) + 0.004s(t)\sum_{i=0}^{29} y(t-i) + 1.5s(t-29)s(t) + 0.201, \notag
\end{eqnarray}
where $s(t)$ are exogenous inputs generated independently and uniformly in the interval $[0,0.5)$. These time series constitute an interesting synthetic example due to the long-term 
dependencies they exhibit.
%------------------------------------------------------------------
%\vspace{0.2cm}
\newline
\noindent \textbf{Cauchy class:} We sampled sequences from a stationary Gaussian process
with correlation function given by $c(x_t,x_{t+h})=(1+ |h|^a)^{-\frac{a}{b}}$  \cite{Gneiting2004}.
%where the parameters adhere to $a \in (0, 2]$ and $b > 0$.
We generated $4$ classes of such time series by the permutation of
parameters $a \in \{0.65,1.95\}$ and $b \in \{0.1,0.95\}$. 
We generated from each class $100$ time series of length $2000$.
%------------------------------------------------------------------

%\vspace{0.005cm}
%\newline
\noindent \textbf{X-ray radiation from black hole binary:} We used data from \cite{Harikrishnan2011} concerning a black hole binary system that expresses various types of temporal regimes which vary over a wide range of time scales.
We extracted subsequences of length $1000$ from regimes $\beta$ and $\kappa$ that
were chosen on account of their similarity (see Fig. \ref{fig:example_regimes}).

%=========================================================================
\section{Discussion and Conclusion}
%=========================================================================

We show the visualisations in Fig. \ref{fig:results}. Unlike t-SNE which operates directly on the raw data,
the proposed algorithm can capture the differences between the time series in the lower dimensional space.
This is because our method explicitly accounts for the sequential nature of time-series;
learning is performed in the space of readout weight representations and is guided by an objective function that
quantifies the reconstruction error in a principled manner.
Of course, the perfectly capable t-SNE is  used here as a mere candidate from the 
class of algorithms designed to visualise vectorial data in order to demonstrate this issue.
Moreover, we demonstrate that our method, by its very nature, is capable of projecting also unseen hold-out data. 
%We  presented a modified autoencoder for visualising time series.
%The presented algorithm relies on representing the time series as readout weights
%of an ESN and endowing the autoencoder with an objective function that
%quantifies the reconstruction error in a principled manner.
% As demonstrated in the
%experiments, the visualisation captures 
%important characteristics of the time series that are otherwise ignored
%when treating the time series merely as vectorial data.
Future work will focus on processing
large datasets of astronomical light curves.

% ****************************************************************************
% BIBLIOGRAPHY AREA
% ****************************************************************************

\begin{footnotesize}

\bibliographystyle{unsrt}
\bibliography{mybiblio}

\end{footnotesize}

% ****************************************************************************
% END OF BIBLIOGRAPHY AREA
% ****************************************************************************

\end{document}